\documentclass[aps,prb,twocolumn,showpacs,superscriptaddress,longbibliography]{revtex4-1}  
\usepackage{graphicx}  
\usepackage{dcolumn}   
\usepackage{bm}        
\usepackage{amssymb}   
\usepackage{multirow}
\usepackage{amsmath}
\usepackage{natbib}
\usepackage{xcolor}
\usepackage[unicode=true,colorlinks=true]{hyperref}
\begin{document}

\title{Optical properties of potential-inserted quantum wells in the near infrared and Terahertz ranges}

\newcommand{\affA}{Laboratoire de Physico-chimie des Microstructures et Micro-syst\`emes, Institut Pr\'eparatoire aux Etudes Scientifiques et Techniques, BP51, 2070 La Marsa, Tunisia}
\newcommand{\affB}{Tyndall National Institute, Lee Maltings, Dyke Parade, Cork, Ireland}
\newcommand{\affC}{ICMN, UMR 7374, Universit\'e d'Orl\'eans and CNRS, Orl\'eans, France}
\newcommand{\affD}{FOTON, Universit\'e Europ\'eenne de Bretagne, INSA-Rennes and CNRS, Rennes, France}
\newcommand{\affE}{Laboratoire de Photonique et Nanostructures, CNRS and Universit\'e Paris-Saclay, route de Nozay, F-91460, Marcoussis, France}

\author{F.~Raouafi}
\affiliation{\affA}
\author{R.~Samti}
\affiliation{\affA}
\author{R.~Benchamekh}
\affiliation{\affB}
\author{R.~Heyd}
\affiliation{\affA}\affiliation{\affC}
\author{S.~Boyer-Richard}
\affiliation{\affD}
\author{P.~Voisin}
\affiliation{\affE}
\author{J-M.~Jancu}
\affiliation{\affD}
\date{\today}
\begin{abstract}
We propose an engineering of the optical properties of GaAs/AlGaAs quantum wells using AlAs and InAs monolayer insertions. A quantitative study of the effects of the monolayer position and the well thickness on the interband and intersubband transitions, based on the extended-basis $sp^3d^5s^* $ tight-binding model, is presented. The effect of insertion on the interband transitions is compared with existing experimental data. As for intersubband transitions, we show that in a GaAs/AlGaAs quantum well including two AlAs and one InAs insertions, a three level \{$e_1$, $e_2$, $e_3$\} system where the transition energy $e_3-e_2$ is lower and the transition energy $e_2-e_1$ larger than the longitudinal optical phonon energy (36 meV) can be engineered together with a $e_3-e_2$ transition energy widely tunable through the TeraHertz range.

\end{abstract}

\maketitle
\section{Introduction}

Semiconductor quantum wells (QW) are the building blocks of a large number of opto-electronic devices. Interband transitions in QWs have been used in devices such QW lasers, modulators and detectors while intersubband (ISB) transitions have allowed engineering of quantum cascade lasers and QW infrared photodetectors\cite{Unuma2011}. Due to unlimited choice in terms of width and material combinations, these structures can be used in opto-electronic applications covering a wide spectral range. Moreover, different techniques to tune the opto-electronic properties of QWs have been proposed such as strain engineering\cite{Adams1986}, doping\cite{Bastard1982,Hawrylak1994}, or monolayer insertion\cite{Marzin1989,Mejri2006,Samti2012}. The latter have been used successfully for tailoring the interband transitions and their Stark shift in electro-absorption modulators\cite{Guettler1997}
by simply changing the position of the insertion within the well. 

More recently, strong effort has been devoted to extending the range of ISB applications to the terahertz (THz) range \cite{Tredicucci2002}. Although asymmetric semiconductor QW structures can be used as laser sources of controllable THz radiation, whatever the active region design, intrinsic non-radiative recombination processes are generally faster than the radiative lifetime of excited states, which hinders the population inversion between subbands\cite{miles2001terahertz}. QWs based on GaAs and AlAs are particularly interesting for intersubband applications thanks to the close lattice constants which guarantees small epitaxial stress and allows the thick epilayers needed in QCL devices. In existing THz QCLs, the lifetime of the lower laser state is determined by the tunneling electron extraction time into the collector/injector and do not display a strong temperature dependence. Conversely, the lifetime of the upper laser state strongly increases when increasing the temperature. This is a major limitation, preventing these QCLs from lasing at high temperature ~\cite{Belkin,Williams,Ferreira,Smet}. The design of new active region where the lifetime of both upper and lower laser states have the same temperature dependency would represent an important step toward THz laser device operating at room temperature.

For a quantitative modeling of such structures, different approaches may be considered. First principle calculations cannot be used, due to the large size of relevant systems (several hundred atom supercells) and their limited precision for excited states representation. Conversely, empirical-parameter models, such as the \textbf{k.p} or envelope-function theory (EFT)\cite{Bastard1988,Smith1990} or atomistic approaches like the empirical pseudopotential\cite{Kim2002} and tight-binding (TB) methods, are suitable for this purpose. While EFT is still the most widely used technique, its applicability to the sub-nm scale (like monolayer insertions) is questionable. Conversely, atomistic approaches have no methodological limitation in handling the case of ultrathin layers and chemical discontinuities. In particular, the TB method and its extended basis $sp^3d^5s^* $ model have been shown to provide a description of the band structure with sub-meV precision throughout the Brillouin zone of the binary semiconductors as well as their heterostructures \cite{Jancu1998,Jancu2004}. 

In this work, we present a quantitative study of the effects of an inserted AlAs monolayer on the interband and intersubband transitions in a GaAs/GaAlAs QW. The results for interband transitions are compared to existing experimental data. A large tunability of transition energies and dipole moments between electron subbands are demonstrated in the THz window as function of the insert position. We also show that a new functionality, the fast depopulation of intermediate state by LO-phonons in a 3-level scheme can be engineered. The paper is organized as follows: in sec.~II, we describe the considered structure and the simulation model. In sec.~III and sec.~IV, we present and discuss the effects of the insert position on the transition energies and the optical dipole momentum of the interband and intersubband transitions respectively, before we conclude in sec.~V.

\section{System of interest and methods}

In the present study, we consider a (001)-oriented (GaAs)$_n$/(AlGaAs)$_m$ superlattice structure where $n$ and $m$ are the numbers of the molecular monolayers (ML) of GaAs and GaAlAs respectively. The basic unperturbed structure studied in this work consists in a GaAs well with $\sim$15 nm width (52 ML), alternating with $\mathrm{Al}_{0.3}\mathrm{Ga}_{0.7}\mathrm{As}$ barriers grown on a GaAs (001) substrate. The barrier is thick enough to ensure that the tunnel coupling between adjacent wells be negligible. The alloyed barrier is modeled using virtual crystal approximation (VCA) where the TB parameters of the virtual crystal are arithmetic means of the constituent material parameters weighted to their concentrations. The unperturbed system is characterized by the D$_{2d}$ point group symmetry and interband transitions obey a parity selection rule with a largely dominant $\delta n = 0$ contribution. Unless perfectly centered, the AlAs monolayer insertion reduces the symmetry level to C$_{2v}$ and breaks the parity selection rule. When the well contains an even number of monolayers, this holds for any position of the insert since the latter will always be surrounded by GaAs layers of slightly different thicknesses, e.g. at best 26 ML and 25 ML respectively for a 52 ML QW. Conversely, for a QW with odd ML number, the insert can be perfectly centered and in this particular case the potential inserted QW retains the D$_{2d}$ point group symmetry. Finally, it is noteworthy that in addition to breaking the parity selection rule, C$_{2v}$ symmetry allows for in-plane optical polarization anisotropy, that is particularly visible in the narrow spectral range between the e$_1$-hh$_1$ and e$_1$-lh$_1$ transitions\cite{Krebs1996,Toropov2000}.

The electronic structure of the superlattices is computed using the extended basis $sp^3d^5s^*$ TB model based on Jancu \textit{et al} parameters\cite{Jancu1998}. This parametrization has shown to allow excellent full-band representation of bulk semiconductors and a good transferability to quantum heterostructures\cite{Jancu1998,Jancu2004}. The supercell is built by translating the standard cubic zinc-blende cell in the $z$ direction. Each cubic cell contains two molecular ML. So, for a (GaAs)$_n$/(AlGaAs)$_m$ superlattice structure we have $(n+m)/2$ cubic cells containing eight atoms each. In the spds* TB model, each atom contributes with twenty orbitals, so the system is represented by a $80(n+m) \times 80(n+m)$ Hamiltonian matrix. For each atom, the $20\times20$ intra-atomic Hamiltonian matrix is an average of the intra-atomic Hamiltonians in the up to four bulk compounds it composes with its four nearest neighbors. The lattice mismatch between AlAs and GaAs is neglected. However, in the case of InAs insertions, a relaxation of the structure is performed using Keating valence force field (VFF) model.~\cite{Keating}. Modeling of a single InAs ML in a GaAs matrix compares satisfactorily with experiments \cite{Brandt1990,Raouafi2016}. We used a GaAs/AlAs and InAs/GaAs valence band offsets (VBOs) of 0.56~eV and 0.4~eV respectively. In order to calculate the optical properties of the superlattices, we have coupled the TB Hamiltonian to an electromagnetic field, following the approach described in refs.~\cite{Ram-Mohan}. This coupling model produces bulk semiconductor dielectric function in good agreement with experimental results and accurately accounts for QW optical absorption spectra, in particular interface-induced optical anisotropy\cite{Krebs1997}.

\section{Interband transitions}
\begin{figure}
\includegraphics[width=\linewidth]{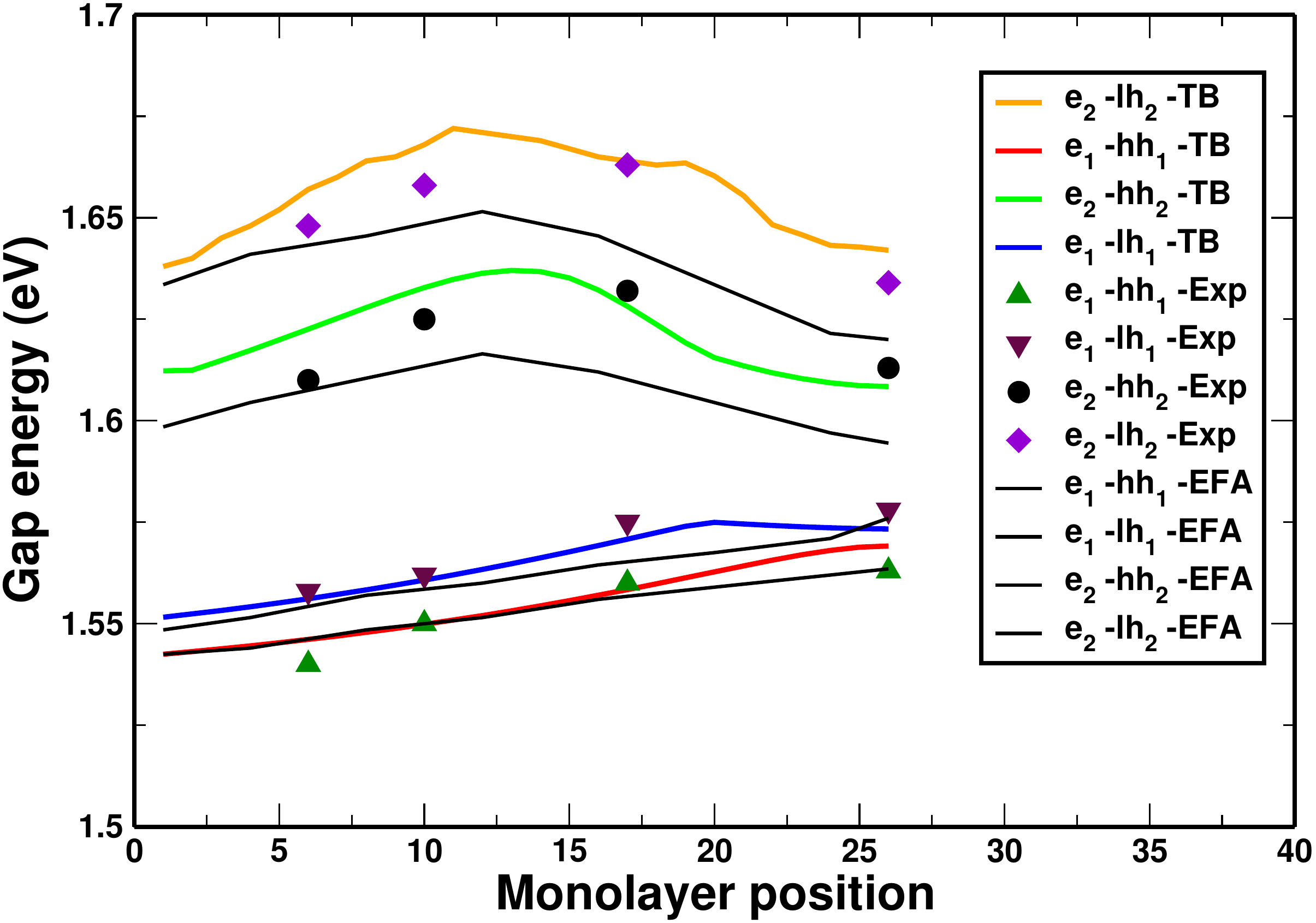}
\caption{Interband energy transitions of a $\sim$15 nm-thick GaAs/GaAlAs multiple quantum wells as a function of the position of the AlAs monolayer from the barrier (in MLs).}
\label{fig1}
\end{figure}

\begin{figure}
\includegraphics[width=\linewidth]{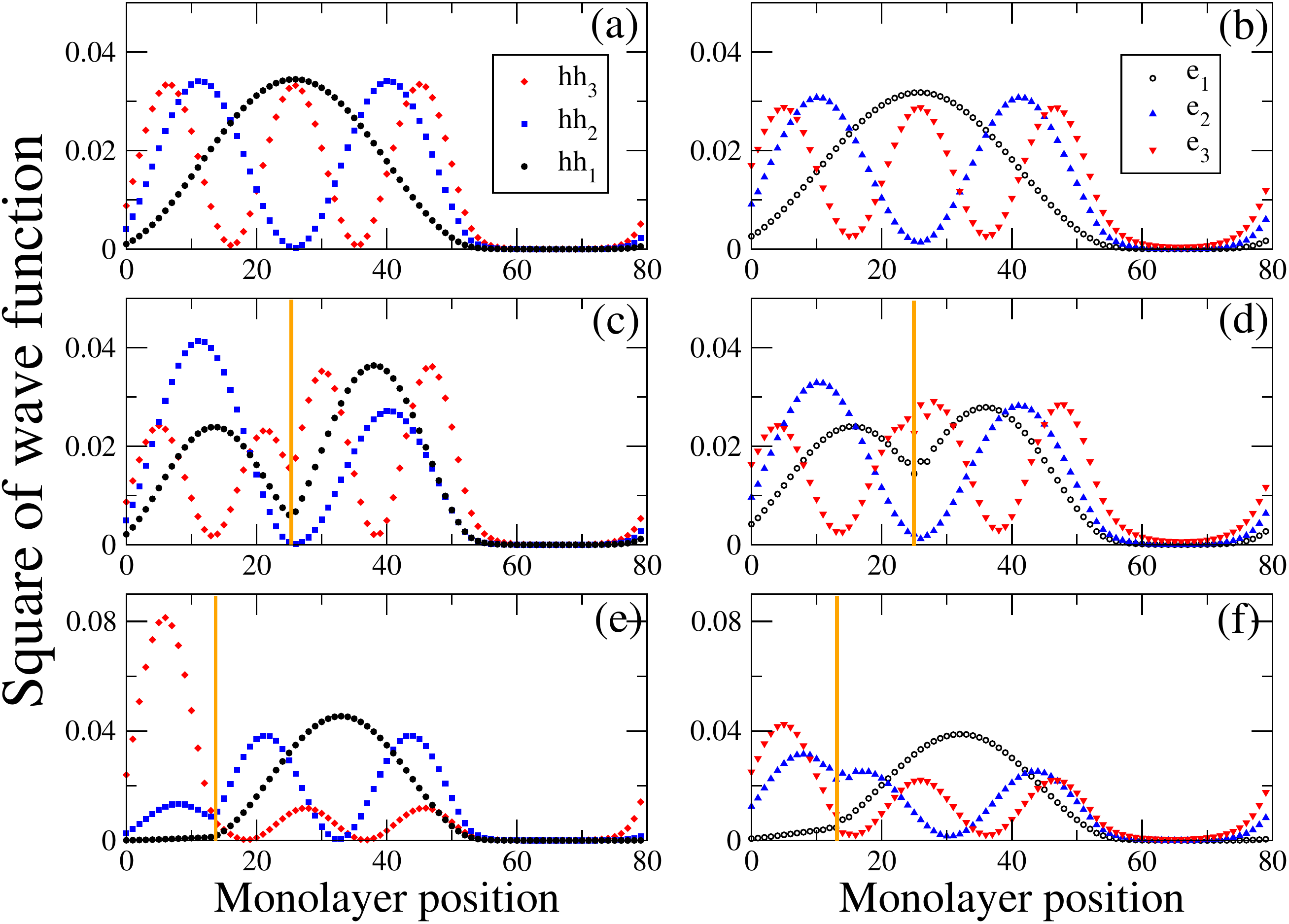}
\caption{Tight-binding results for the hole (left) and electron (right) state squared wave function of a $\sim$15 nm-thick GaAs/GaAlAs quantum well: without AlAs insertion (a) and (b) , with the AlAs ML inserted at mid-well (c) and (d) and  at $\sim$3.75 nm away from the barrier (e) and (f).}
\label{fig2}
\end{figure}

We firstly focus on the interband transitions energies and the corresponding dipole matrix elements. Calculations of electronic band structure of the systems described in the previous section are performed with varying the AlAs monolayer position from the left barrier to the center of the well. Figure \ref{fig1} shows the optical transitions energies calculated using our tight binding model, compared with envelope function approximation (EFA) and experimental data\cite{Marzin1989,Mejri2006} obtained from low temperature photoluminescence, excitation and photo-reflectance spectra.
The transition energies were calculated at the center of the Brillouin zone ($\Gamma$-point) and labelled according to the dominant bulk-state component of initial and final states: conduction (e), heavy-hole (hh) and light-hole (lh). As can be seen from the figure \ref{fig1}, the TB calculation agrees with experiment within experimental uncertainties. In particular, the predicted reduction of the hh1-lh1 splitting for an insert at the mid-well is clearly observed in PLE data of Ref.~\citenum{Marzin1989}. The EFA results \cite{Mejri2006} are also given in the figure for comparison. Despite the still reasonable agreement of these results with experience, one should keep in mind that there is no theoretical ground for modeling an AlAs ML insertion by 3~\AA~ of bulk AlAs material, using standard EFT continuity relations.
It can be observed that the blueshift of the e$_1$-hh$_1$ and e$_1$-lh$_1$ transitions is maximal when the ML is localized at mid-well ($\sim$7.5 nm or 26 MLs). In contrast, the maxima of the e$_2$-hh$_2$ and e$_2$-lh$_2$ transitions occur when the insert is located at about the quarter of the well ($\sim$3.75 nm). A qualitative explanation of these observations can be obtained from a first-order perturbation approach where the insert act as a delta-like repulsive perturbation on the original QW states. Indeed, as it can be seen on the figure~\ref{fig2}(a), the largest amplitudes of the unperturbed e$_1$, lh$_1$ and hh$_1$ ground-state wave functions are located at the mid-well, whereas the maxima corresponding to e$_2$ and hh$_2$ are located at the quarter of the QW. When the insert is localized at mid-well (Fig.~\ref{fig2}(b)), the effect of the perturbation is important for hh$_1$, lh$_1$ and e$_1$ and nearly zero for hh$_2$, lh$_2$ and e$_2$. Conversely, when the AlAs ML is located at the quarter of the well ($\sim$3.75 nm, Fig. \ref{fig2}(c)), all the states (hh$_1$, lh$_1$, e$_1$, e$_2$ and hh$_2$) are strongly affected by the perturbing potential. Yet, the calculated charge densities (Fig.~\ref{fig2}(b, c)) evidence that the AlAs monolayer act as a strong perturbation that significantly affects the wavefunctions, hence the perturbative framework is unable to account quantitatively for the results. In this sense, the AlAs ML cannot be considered as a ``probe'' of the unperturbed QW eigenstates.

\par

\par%

\begin{figure}
\centering
\includegraphics[width=\linewidth]{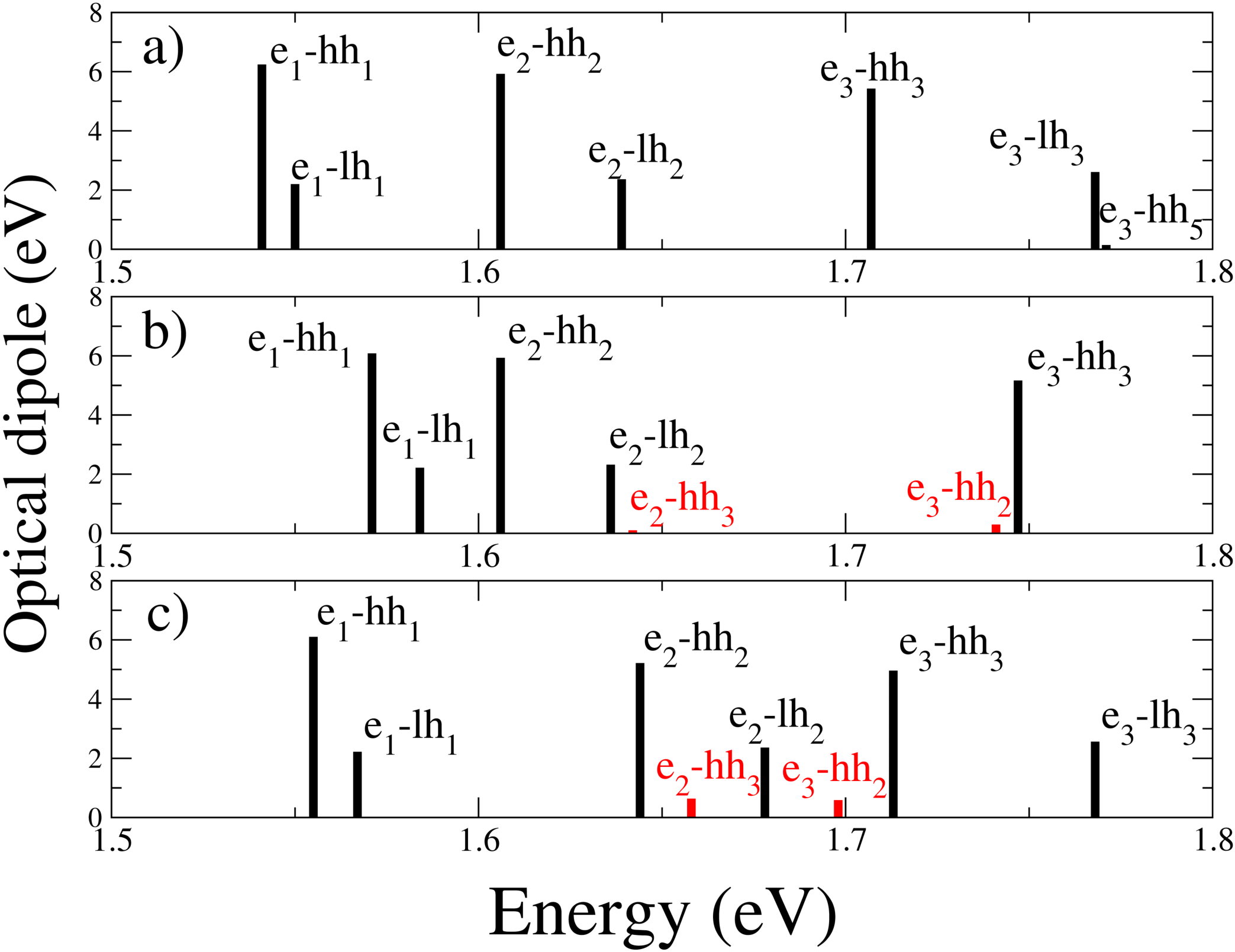} \includegraphics[width=\linewidth]{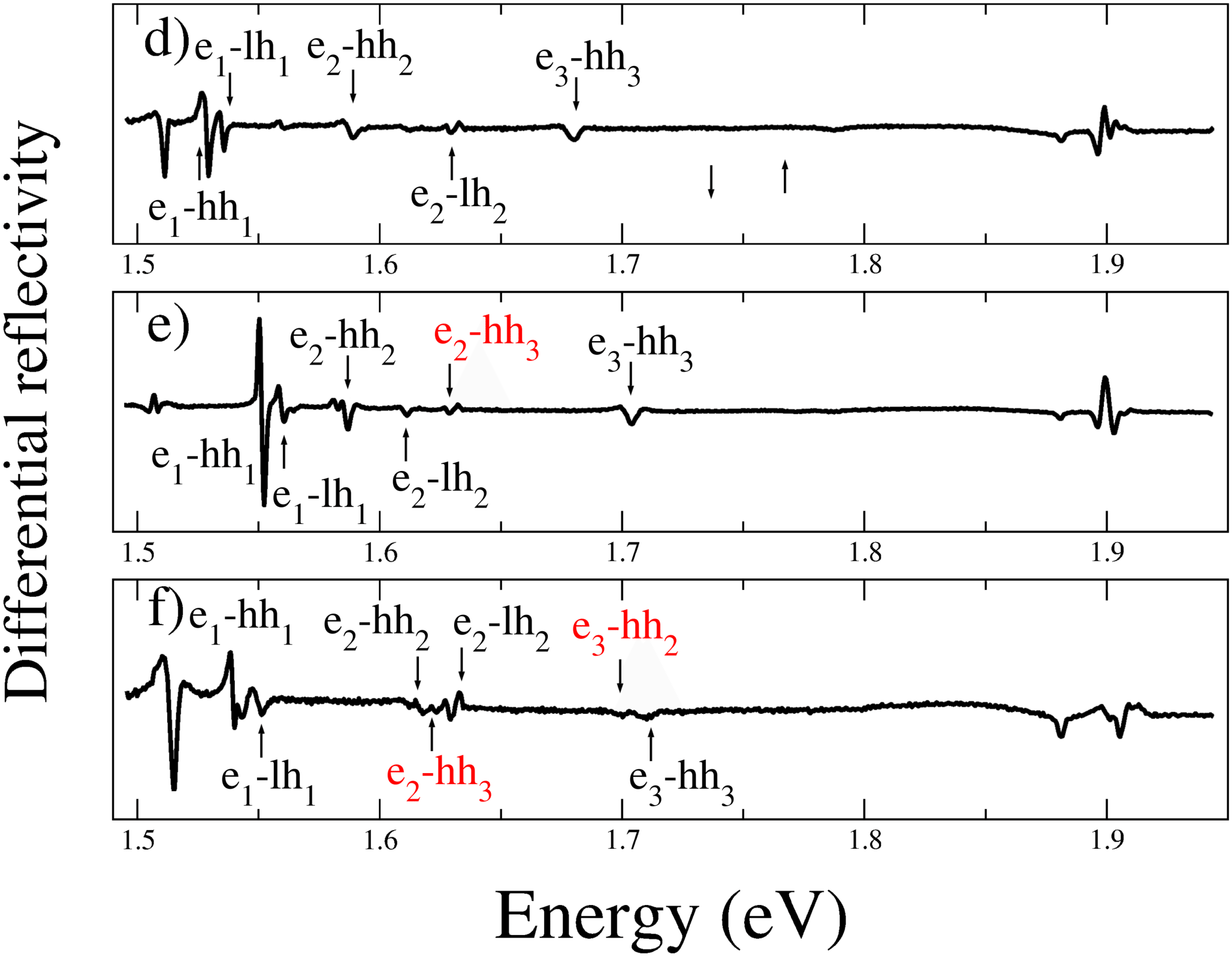}
\caption{Calculated squared dipole matrix elements (E$_p$) for transverse electric polarization between valence and conduction subbands (top) and differential reflectivity measurements (bottom) of a $\sim$15 nm-thick GaAs/GaAlAs quantum well: without AlAs insertion (a) and (d), with the AlAs ML inserted at mid-well (b) and (e) and  at $\sim$3.75 nm away from the barrier (c) and (f). Experimental features at 1.520~eV and 1.9~eV correspond respectively to GaAs substrate and  Al$_{0.3}$Ga$_{0.7}$As barrier}
\label{fig3}
\end{figure}

Next, we calculate the dipole matrix elements corresponding to the interband transitions. Figure~\ref{fig3} shows the squared dipole transition elements ($E_P$ in eV) for three different samples: (a) the unperturbed well, (b) the insert placed at the center of the well and (c) the insert placed at $\sim$3.75nm from the barrier. In the framework of the envelope-function theory \cite{Bastard1988}, the optical transitions for a symmetric QW (like the sample (a)) obey a parity selection rule ($\Delta n$ even), with a strongly dominant $\Delta n =0$ contribution. The results shown in Fig.~\ref{fig3}(a) clearly obey this selection rule. However, the insertion of the AlAs ML breaks the symmetry and thus several parity-forbidden transitions appear in the spectrum, such as e$_{1}$-hh$_{2}$,e$_{2}$-lh$_{1}$, e$_{1}$-lh$_{2}$, e$_{2}$-hh$_{3}$ and e$_{3}$-hh$_{2}$. Only the last two transitions actually have enough oscillator strength to be visible in Fig.~\ref{fig3} (b, c). This symmetry breaking occurs even for the sample (b) where the insert is placed at the 26$^{th}$ ML. Indeed, as stated in sec.~II, since the well contains an even number of MLs, replacing the 26$^{th}$ GaAs ML by an AlAs ML create a slightly non-symmetric structure. This also can be confirmed by the results in Fig.~\ref{fig2}(b) where the wave functions of ground hole and electron states are not symmetric in respect to the AlAs ML. Yet, despite the huge asymmetry of the wavefunctions in the potential-inserted quantum well, the parity-forbidden $\Delta n =1$ transitions remain quite weak and the parity-allowed transitions nearly retain their original oscillator strength. This reflects the fact that, unlike in the situation of a biased QW, the conduction and valence band potentials remain ``homomorph'', whatever the position of the insert is: valence and conduction eigenstates become strongly asymmetric, but they remain similar to each other. Since different eigenstates within a given band are necessarily orthogonal,  the interband overlaps still obey an approximate $\Delta n =0$ selection rule.

\par
In order to validate our modeling, we have compared our computational results to those provided by photoreflectance (PR) spectroscopy \cite{Mejri2006}, which is a well known technique commonly used to investigate the optical properties of QWs over a large energy range. The multi quantum-well structures were grown by a conventional solid-source molecular beam epitaxy \cite{Marzin1989}. The samples consist of sequences of six $\sim$15 nm thick GaAs wells separated by a $\sim$8.5 nm thick Al$_{0.3}$Ga$_{0.7}$As barrier. A single AlAs monolayer has been inserted at different positions inside the well. The optical transitions energies were determined  at low temperature (10 K), from the reflectivity measurements. Figure~\ref{fig3}(d) shows the differential reflectivity data obtained in the case of the reference sample (without monolayer insertion) and in the case of the samples containing one AlAs monolayer inserted inside the well, respectively at the middle (Fig.~\ref{fig3}(e)) and at the quarter (Fig.~\ref{fig3}(f)) of the QW. The experimental spectra are dominated by the easily identified e$_1$-hh$_1$ excitonic transition and all the other peaks can be precisely assigned by comparison with the present modeling.

\section{Intersubband transitions}

\begin{figure}
\includegraphics[width=0.48\linewidth]{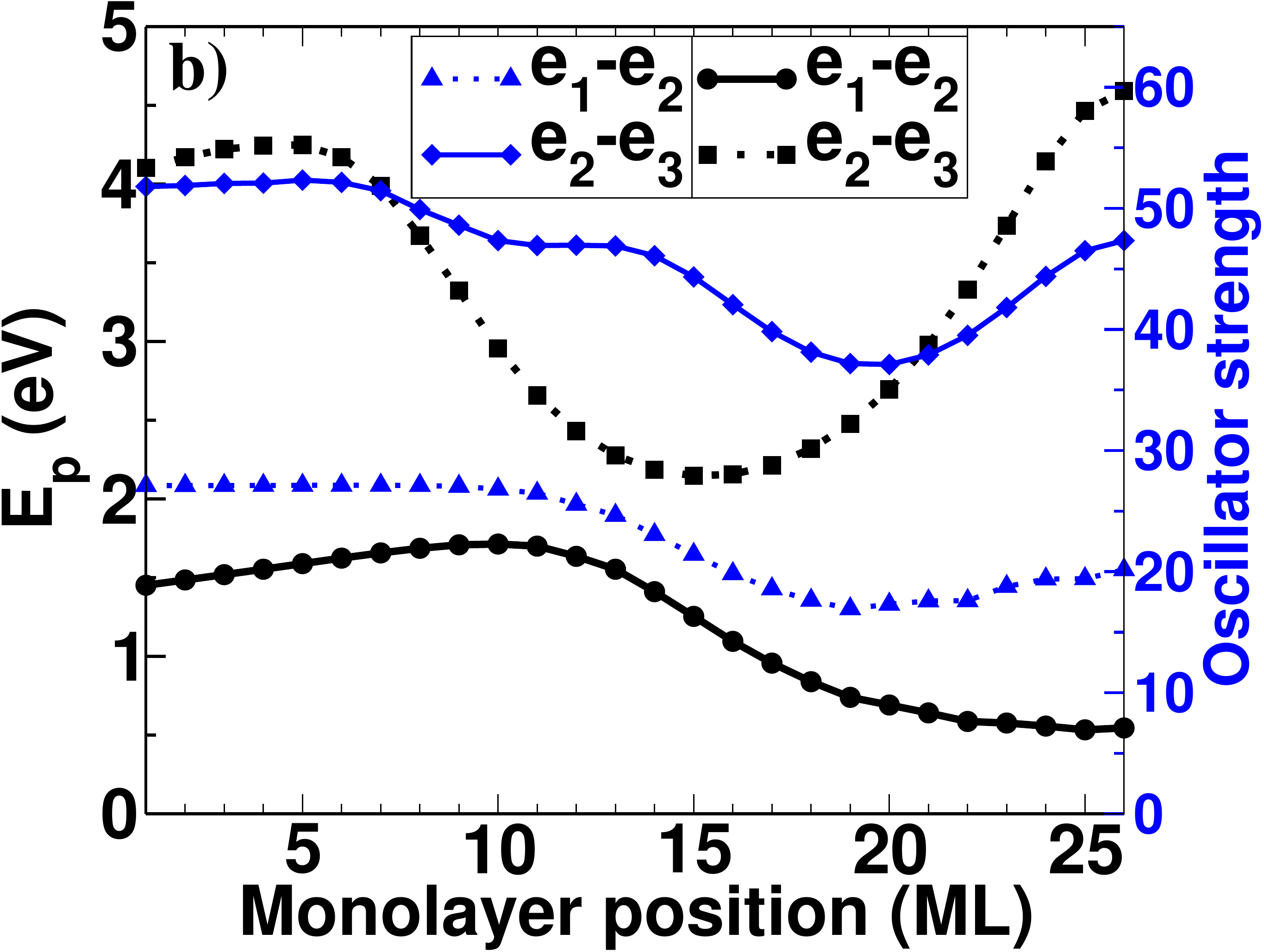}\quad \includegraphics[width=0.48\linewidth]{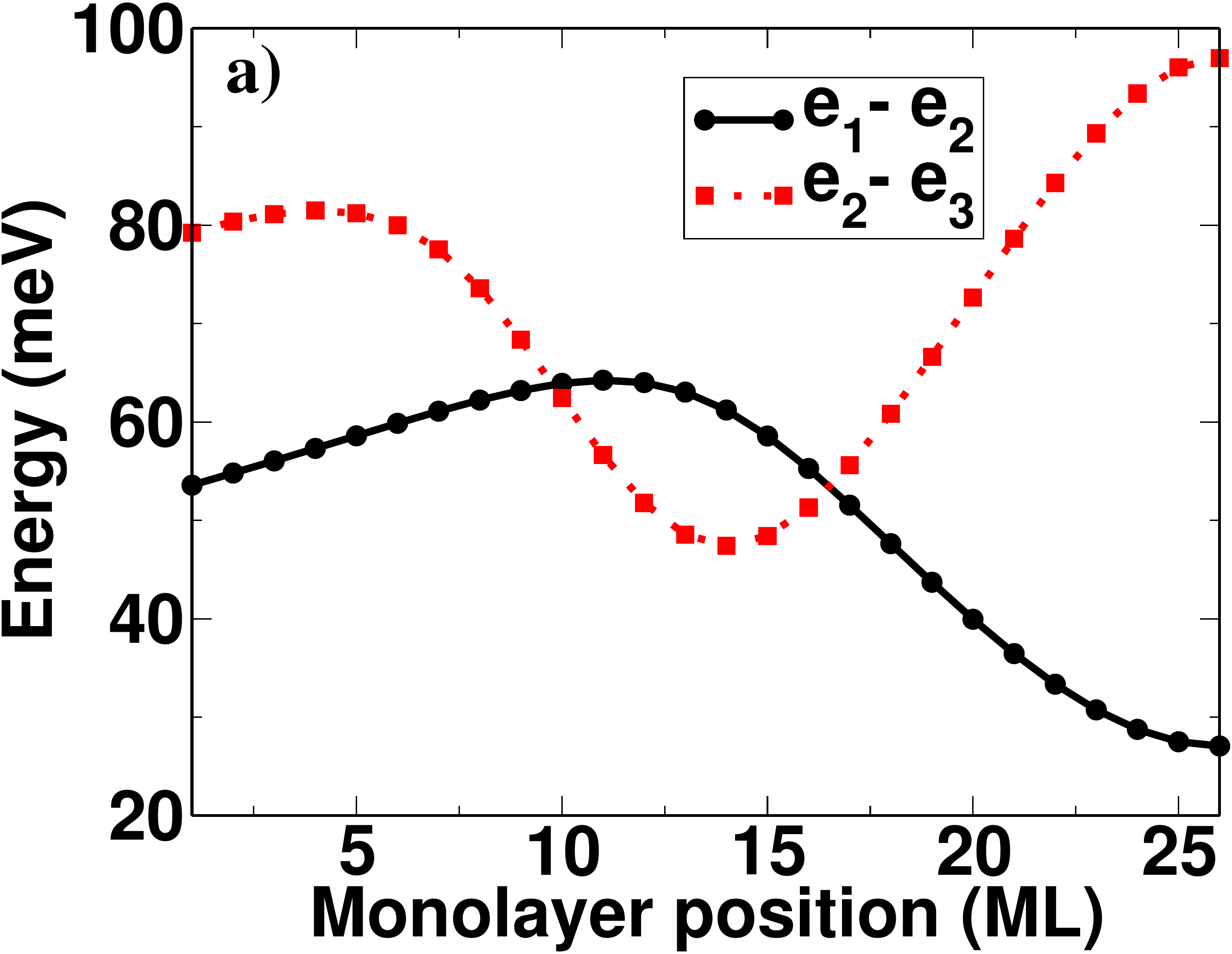}
\caption{Dependence of the intersubband transition energies e$_1$-e$_2$ and e$_2$-e$_3$ on the AlAs monolayer position for a $\sim$6.8 nm (left) and $\sim$15 nm-thick (right) GaAs/GaAlAs quantum well .}
\label{fig5}
\end{figure}

The intersubband transitions in GaAs/GaAlAs QWs have attracted much attention during the last decades, due to their use as active region of quantum cascade lasers and infrared detectors. Sources of terahertz (THz) emitters are investigated for their various applications in imaging and wireless communications. The lasers can operate on a three-level model to realize population inversions between the first and second excited states. Our TB  model was demonstrated to give an accurate modeling of intersubband properties (dipole moment and energy) in strained quantum wells~\cite{Jancu98,Jancu2000}. After corroborating our results for AlAs and InAs MLs insertion by comparison to experimental results of interband transitions in the previous section and in Ref.~\citenum{Samti2012}, respectively, we investigate here the possibility of engineering a three level system {e$_1$, e$_2$, e$_3$} where the e$_2$-e$_3$ transition energy is lower and  the e$_1$-e$_2$ transition energy larger than the LO phonon energy (36 meV). In a first step, we consider the effects of one AlAs ML insertion on the intersubband transition energies and optical matrix element of a 52 ML thick GaAs/Ga$_{0.7}$Al$_{0.3}$As QW (see Fig.~\ref{fig5}.a). We note that the e$_1$-e$_2$ transition energy is minimum when the ML is at the mid-well, then it increases to reach a maximum value when the ML is at a certain position between the barrier and the mid-well before it decreases again. In contrast, the  e$_2$-e$_3$ transition energy is maximum when the AlAs ML is at the mid-well and has a minimum value for a ML position located between the barrier and the mid-well. This can be explained by a perturbative analysis of the wavefunctions of e$_1$, e$_2$ and e$_3$ states of the unperturbed QW (cf. Fig.~\ref{fig2}.b). Indeed, the AlAs ML insertion acts like a repulsive potential that shifts up all electron state energies. However, since at the mid-well both e$_1$ and e$_3$ wave functions are maximal while e$_2$ wave function is minimal, the ML shifts up the e$_1$ and e$_3$ energies more than the e$_2$ energy, and thus increases the e$_2$-e$_3$ transition energy and reduces the e$_1$-e$_2$ transition energy with respect to the unperturbed QW transition energies. Similarly, when the ML is inserted close to the third of the QW, it has a higher effect on the e$_2$ state energy and a negligible effect on the e$_3$ state energy. The later configuration, {\it i. e.} when the ML is at the 14$^{th}$ ML, is more interesting since the e$_2$-e$_3$ transition energy is lower than the e$_1$-e$_2$ one. However, the e$_2$-e$_3$ transition energy is still larger than the LO phonon energy, and so further engineering (see below) is needed. Figure~\ref{fig5}.b shows the dependence of the optical matrix element and the oscillator strength of the e$_1$-e$_2$ and e$_2$-e$_3$ transitions on the insert position. It is clear that both transitions are still optically active wherever the ML insertion is placed. Moreover, the oscillator strengths of both transitions is not drastically affected by the insertion. These results are encouraging to perform further engineering of the optical properties through the insertion of multiple MLs. 
\par

In order to obtain a smaller value for the e$_2$-e$_3$ transition energy, we consider a 52 ML thick GaAs/Ga$_{0.7}$Al$_{0.3}$As QW with two AlAs ML insertion. Inspired with the previous results the two ML are placed at the 14$^{th}$ and the 39$^{th}$ ML, respectively. The transition energies calculated for this structure are: e$_1$-e$_2$=63~meV and e$_2$-e$_3$=40~meV, while the corresponding oscillator strength are 24.70 and 41.64, respectively. The e$_2$-e$_3$ transition energy is closer to the LO phonon energy, but still larger. Further engineering can be done by inserting an extra InAs ML near the QW center. Indeed, the InAs ML acts like an attractive potential and so reduces the e$_1$ and e$_3$ energies while leaving e$_2$ unchanged. Hence the InAs ML insertion further reduces the e$_2$-e$_3$ transition energy and increases the e$_1$-e$_2$ transition energy. 

Next, we consider a 52 ML thick GaAs/Ga$_{0.7}$Al$_{0.3}$As QW with two AlAs and one InAs MLs inserted at the 14$^{th}$, the 39$^{th}$ and the 26$^{th}$ MLs respectively. The calculated transition energies are: e$_1$-e$_2$=125~meV and e$_2$-e$_3$=25~meV, while the corresponding oscillator strength are 17.68 and 32.99, respectively. This structure is very promising to achieve THz generation at room temperature since the e$_2$-e$_3$ transition energy is lower and the e$_1$-e$_2$ transition energy is higher than the LO phonon energy, while the oscillator strength of the e$_2$-e$_3$ transition is only 34\% smaller than the oscillator strength of the corresponding transition in the unperturbed QW. Furthermore, the large value of the e$_1$-e$_2$ transition energy allows for further tuning of the structure by changing the QW thickness, the barrier composition or the MLs position.

Finally, we investigate the possibility of tuning the e$_1$-e$_2$ transition energy by modifying the well thickness and the ML positions. Table~\ref{tab1} summarize the results for a selection of different structure configurations. We show here that the e$_2$-e$_3$ transition energy can be tuned to cover all the energies below the LO phonon energy and its oscillator strength stay in the same order of magnitude as in the unperturbed QW while the e$_1$-e$_2$ transition energy is kept far above 36meV. These results indicate that the insert-based engineering technique is promising for the realization of THz laser at room temperature.

\begin{table}\caption{Transition energy and oscillator strength of e$_1$-e$_2$ and e$_2$-e$_3$ transition in GaAs/Ga$_{0.7}$Al$_{0.3}$As QWs. Here l is the well thickness (in MLs), m and n are the positions of the two AlAs ML insertions and p is the position of the InAs ML insertion.\label{tab1}}
\begin{tabular*}{\columnwidth}{@{\extracolsep{\fill}}cccccccc}
\hline
\hline
l & m & n & p & \multicolumn{2}{c}{Transition energy (meV)} &  \multicolumn{2}{c}{Oscillator strength}\\
  &   &   &   & e$_1$-e$_2$ & e$_2$-e$_3$   & e$_1$-e$_2$ & e$_2$-e$_3$ \\
\hline
44 & 12 & 33 & 22 & 141 & 34 & 18.74 & 33.37 \\ 
48 & 13 & 36 & 24 & 132 & 29 & 18.22 & 33.13 \\
52 & 14 & 39 & 26 & 125 & 25 & 17.68 & 32.99 \\
56 & 15 & 42 & 28 & 117 & 22 & 17.12 & 32.60 \\
60 & 16 & 45 & 30 & 112 & 19 & 16.57 & 32.31 \\
64 & 17 & 48 & 32 & 106 & 17 & 16.02 & 31.99 \\
68 & 18 & 51 & 34 & 102 & 15 & 15.49 & 31.66 \\
76 & 20 & 57 & 38 &  93 & 12 & 14.47 & 30.97 \\
84 & 22 & 63 & 42 &  86 & 10 & 13.51 & 30.26 \\
92 & 24 & 69 & 46 &  80 &  8 & 12.61 & 29.55 \\
108& 28 & 81 & 54 &  72 &  5 & 11.06 & 28.18 \\ 
\hline
\hline
\end{tabular*}
\end{table}

\section{Conclusion}
We have studied the effects of AlAs monolayer insertion inside a GaAs/GaAlAs quantum well and how the variation of the monolayer position and the well width affect the interband and intersubband transition properties by mean of the $sp^3d^5s^*$ tight-binding model. Firstly, we validated the theoretical results of the interband transitions by comparing to the differential reflectivity measurement. Then we investigated the intersubband transition properties and the possibility of a three electron states system where the upper transition energy is smaller and the lower transition energy larger than the LO phonon energy. We demonstrated that this kind of system can be obtained by three ML insertions, two AlAs and one InAs MLs. We also show that the upper transition energy can be widely tuned below the LO phonon energy while the lower transition is always far above 36~meV.


\bibliography{GaAs_AlAs_AlGaAs}

\begin{thebibliography}{28}%
\makeatletter
\providecommand \@ifxundefined [1]{%
 \@ifx{#1\undefined}
}%
\providecommand \@ifnum [1]{%
 \ifnum #1\expandafter \@firstoftwo
 \else \expandafter \@secondoftwo
 \fi
}%
\providecommand \@ifx [1]{%
 \ifx #1\expandafter \@firstoftwo
 \else \expandafter \@secondoftwo
 \fi
}%
\providecommand \natexlab [1]{#1}%
\providecommand \enquote  [1]{``#1''}%
\providecommand \bibnamefont  [1]{#1}%
\providecommand \bibfnamefont [1]{#1}%
\providecommand \citenamefont [1]{#1}%
\providecommand \href@noop [0]{\@secondoftwo}%
\providecommand \href [0]{\begingroup \@sanitize@url \@href}%
\providecommand \@href[1]{\@@startlink{#1}\@@href}%
\providecommand \@@href[1]{\endgroup#1\@@endlink}%
\providecommand \@sanitize@url [0]{\catcode `\\12\catcode `\$12\catcode
  `\&12\catcode `\#12\catcode `\^12\catcode `\_12\catcode `\%12\relax}%
\providecommand \@@startlink[1]{}%
\providecommand \@@endlink[0]{}%
\providecommand \url  [0]{\begingroup\@sanitize@url \@url }%
\providecommand \@url [1]{\endgroup\@href {#1}{\urlprefix }}%
\providecommand \urlprefix  [0]{URL }%
\providecommand \Eprint [0]{\href }%
\providecommand \doibase [0]{http://dx.doi.org/}%
\providecommand \selectlanguage [0]{\@gobble}%
\providecommand \bibinfo  [0]{\@secondoftwo}%
\providecommand \bibfield  [0]{\@secondoftwo}%
\providecommand \translation [1]{[#1]}%
\providecommand \BibitemOpen [0]{}%
\providecommand \bibitemStop [0]{}%
\providecommand \bibitemNoStop [0]{.\EOS\space}%
\providecommand \EOS [0]{\spacefactor3000\relax}%
\providecommand \BibitemShut  [1]{\csname bibitem#1\endcsname}%
\let\auto@bib@innerbib\@empty
\bibitem [{\citenamefont {Unuma}\ \emph {et~al.}(2011)\citenamefont {Unuma},
  \citenamefont {Takata}, \citenamefont {Sakasegawa}, \citenamefont
  {Hirakawa},\ and\ \citenamefont {Nakamura}}]{Unuma2011}%
  \BibitemOpen
  \bibfield  {author} {\bibinfo {author} {\bibfnamefont {T.}~\bibnamefont
  {Unuma}}, \bibinfo {author} {\bibfnamefont {S.}~\bibnamefont {Takata}},
  \bibinfo {author} {\bibfnamefont {Y.}~\bibnamefont {Sakasegawa}}, \bibinfo
  {author} {\bibfnamefont {K.}~\bibnamefont {Hirakawa}}, \ and\ \bibinfo
  {author} {\bibfnamefont {A.}~\bibnamefont {Nakamura}},\ }\href {\doibase
  http://dx.doi.org/10.1063/1.3549126} {\bibfield  {journal} {\bibinfo
  {journal} {Journal of Applied Physics}\ }\textbf {\bibinfo {volume} {109}},\
  \bibinfo {pages} {043506} (\bibinfo {year} {2011})}\BibitemShut {NoStop}%
\bibitem [{\citenamefont {Adams}(1986)}]{Adams1986}%
  \BibitemOpen
  \bibfield  {author} {\bibinfo {author} {\bibfnamefont {A.}~\bibnamefont
  {Adams}},\ }\href {\doibase 10.1049/el:19860171} {\bibfield  {journal}
  {\bibinfo  {journal} {Electronics Letters}\ }\textbf {\bibinfo {volume}
  {22}},\ \bibinfo {pages} {249} (\bibinfo {year} {1986})}\BibitemShut
  {NoStop}%
\bibitem [{\citenamefont {Bastard}(1982)}]{Bastard1982}%
  \BibitemOpen
  \bibfield  {author} {\bibinfo {author} {\bibfnamefont {G.}~\bibnamefont
  {Bastard}},\ }\href {\doibase http://dx.doi.org/10.1016/0039-6028(82)90580-5}
  {\bibfield  {journal} {\bibinfo  {journal} {Surface Science}\ }\textbf
  {\bibinfo {volume} {113}},\ \bibinfo {pages} {165 } (\bibinfo {year}
  {1982})}\BibitemShut {NoStop}%
\bibitem [{\citenamefont {Hawrylak}\ \emph {et~al.}(1994)\citenamefont
  {Hawrylak}, \citenamefont {Young},\ and\ \citenamefont
  {Brockmann}}]{Hawrylak1994}%
  \BibitemOpen
  \bibfield  {author} {\bibinfo {author} {\bibfnamefont {P.}~\bibnamefont
  {Hawrylak}}, \bibinfo {author} {\bibfnamefont {J.~F.}\ \bibnamefont {Young}},
  \ and\ \bibinfo {author} {\bibfnamefont {P.}~\bibnamefont {Brockmann}},\
  }\href {http://stacks.iop.org/0268-1242/9/i=5S/a=007} {\bibfield  {journal}
  {\bibinfo  {journal} {Semiconductor Science and Technology}\ }\textbf
  {\bibinfo {volume} {9}},\ \bibinfo {pages} {432} (\bibinfo {year}
  {1994})}\BibitemShut {NoStop}%
\bibitem [{\citenamefont {Marzin}\ and\ \citenamefont
  {G\'erard}(1989)}]{Marzin1989}%
  \BibitemOpen
  \bibfield  {author} {\bibinfo {author} {\bibfnamefont {J.-Y.}\ \bibnamefont
  {Marzin}}\ and\ \bibinfo {author} {\bibfnamefont {J.-M.}\ \bibnamefont
  {G\'erard}},\ }\href {\doibase 10.1103/PhysRevLett.62.2172} {\bibfield
  {journal} {\bibinfo  {journal} {Phys. Rev. Lett.}\ }\textbf {\bibinfo
  {volume} {62}},\ \bibinfo {pages} {2172} (\bibinfo {year}
  {1989})}\BibitemShut {NoStop}%
\bibitem [{\citenamefont {Mejri}\ \emph {et~al.}(2006)\citenamefont {Mejri},
  \citenamefont {Chaouache}, \citenamefont {Maaref}, \citenamefont {Voisin},\
  and\ \citenamefont {Gerard}}]{Mejri2006}%
  \BibitemOpen
  \bibfield  {author} {\bibinfo {author} {\bibfnamefont {C.}~\bibnamefont
  {Mejri}}, \bibinfo {author} {\bibfnamefont {M.}~\bibnamefont {Chaouache}},
  \bibinfo {author} {\bibfnamefont {M.}~\bibnamefont {Maaref}}, \bibinfo
  {author} {\bibfnamefont {P.}~\bibnamefont {Voisin}}, \ and\ \bibinfo {author}
  {\bibfnamefont {J.-M.}\ \bibnamefont {Gerard}},\ }\href
  {http://stacks.iop.org/0268-1242/21/i=8/a=005} {\bibfield  {journal}
  {\bibinfo  {journal} {Semiconductor Science and Technology}\ }\textbf
  {\bibinfo {volume} {21}},\ \bibinfo {pages} {1018} (\bibinfo {year}
  {2006})}\BibitemShut {NoStop}%
\bibitem [{\citenamefont {Samti}\ \emph {et~al.}(2012)\citenamefont {Samti},
  \citenamefont {Raouafi}, \citenamefont {Chaouach}, \citenamefont {Maaref},
  \citenamefont {Sakri}, \citenamefont {Even}, \citenamefont {Gerard},\ and\
  \citenamefont {Jancu}}]{Samti2012}%
  \BibitemOpen
  \bibfield  {author} {\bibinfo {author} {\bibfnamefont {R.}~\bibnamefont
  {Samti}}, \bibinfo {author} {\bibfnamefont {F.}~\bibnamefont {Raouafi}},
  \bibinfo {author} {\bibfnamefont {M.}~\bibnamefont {Chaouach}}, \bibinfo
  {author} {\bibfnamefont {M.}~\bibnamefont {Maaref}}, \bibinfo {author}
  {\bibfnamefont {A.}~\bibnamefont {Sakri}}, \bibinfo {author} {\bibfnamefont
  {J.}~\bibnamefont {Even}}, \bibinfo {author} {\bibfnamefont {J.-M.}\
  \bibnamefont {Gerard}}, \ and\ \bibinfo {author} {\bibfnamefont {J.-M.}\
  \bibnamefont {Jancu}},\ }\href {\doibase http://dx.doi.org/10.1063/1.4731783}
  {\bibfield  {journal} {\bibinfo  {journal} {Applied Physics Letters}\
  }\textbf {\bibinfo {volume} {101}},\ \bibinfo {eid} {012105} (\bibinfo {year}
  {2012})}\BibitemShut {NoStop}%
\bibitem [{\citenamefont {Guettler}\ \emph {et~al.}(1997)\citenamefont
  {Guettler}, \citenamefont {Krebs}, \citenamefont {Dias}, \citenamefont
  {Harmand}, \citenamefont {Devaux},\ and\ \citenamefont
  {Voisin}}]{Guettler1997}%
  \BibitemOpen
  \bibfield  {author} {\bibinfo {author} {\bibfnamefont {T.}~\bibnamefont
  {Guettler}}, \bibinfo {author} {\bibfnamefont {O.}~\bibnamefont {Krebs}},
  \bibinfo {author} {\bibfnamefont {I.~L.}\ \bibnamefont {Dias}}, \bibinfo
  {author} {\bibfnamefont {J.~C.}\ \bibnamefont {Harmand}}, \bibinfo {author}
  {\bibfnamefont {F.}~\bibnamefont {Devaux}}, \ and\ \bibinfo {author}
  {\bibfnamefont {P.}~\bibnamefont {Voisin}},\ }\href
  {http://stacks.iop.org/0268-1242/12/i=6/a=014} {\bibfield  {journal}
  {\bibinfo  {journal} {Semiconductor Science and Technology}\ }\textbf
  {\bibinfo {volume} {12}},\ \bibinfo {pages} {729} (\bibinfo {year}
  {1997})}\BibitemShut {NoStop}%
\bibitem [{\citenamefont {Kohler}\ \emph {et~al.}(2002)\citenamefont {Kohler},
  \citenamefont {Tredicucci}, \citenamefont {Beltram}, \citenamefont {Beere},
  \citenamefont {Linfield}, \citenamefont {Giles}, \citenamefont {Ritchie},
  \citenamefont {Iotti},\ and\ \citenamefont {Rossi}}]{Tredicucci2002}%
  \BibitemOpen
  \bibfield  {author} {\bibinfo {author} {\bibfnamefont {R.}~\bibnamefont
  {Kohler}}, \bibinfo {author} {\bibfnamefont {A.}~\bibnamefont {Tredicucci}},
  \bibinfo {author} {\bibfnamefont {F.}~\bibnamefont {Beltram}}, \bibinfo
  {author} {\bibfnamefont {H.~E.}\ \bibnamefont {Beere}}, \bibinfo {author}
  {\bibfnamefont {E.~H.}\ \bibnamefont {Linfield}}, \bibinfo {author}
  {\bibfnamefont {D.~A.}\ \bibnamefont {Giles}}, \bibinfo {author}
  {\bibfnamefont {D.~A.}\ \bibnamefont {Ritchie}}, \bibinfo {author}
  {\bibfnamefont {R.~C.}\ \bibnamefont {Iotti}}, \ and\ \bibinfo {author}
  {\bibfnamefont {F.}~\bibnamefont {Rossi}},\ }\href {\doibase
  http://dx.doi.org/10.1038/417156a} {\bibfield  {journal} {\bibinfo  {journal}
  {Nature}\ }\textbf {\bibinfo {volume} {417}},\ \bibinfo {pages} {156–159}
  (\bibinfo {year} {2002})},\ \bibinfo {note} {10.1038/417156a}\BibitemShut
  {NoStop}%
\bibitem [{\citenamefont {Miles}\ \emph {et~al.}(2001)\citenamefont {Miles},
  \citenamefont {Harrison},\ and\ \citenamefont
  {Lippens}}]{miles2001terahertz}%
  \BibitemOpen
  \bibfield  {author} {\bibinfo {author} {\bibfnamefont {R.}~\bibnamefont
  {Miles}}, \bibinfo {author} {\bibfnamefont {P.}~\bibnamefont {Harrison}}, \
  and\ \bibinfo {author} {\bibfnamefont {D.}~\bibnamefont {Lippens}},\
  }\href@noop {} {\emph {\bibinfo {title} {Terahertz sources and systems}}},\
  Vol.~\bibinfo {volume} {27}\ (\bibinfo  {publisher} {Springer Science \&
  Business Media},\ \bibinfo {year} {2001})\BibitemShut {NoStop}%
\bibitem [{\citenamefont {Belkin}\ \emph {et~al.}(2009)\citenamefont {Belkin},
  \citenamefont {Wang}, \citenamefont {Pflugl}, \citenamefont {Belyanin},
  \citenamefont {Khanna}, \citenamefont {Davies}, \citenamefont {Linfield},\
  and\ \citenamefont {Capasso}}]{Belkin}%
  \BibitemOpen
  \bibfield  {author} {\bibinfo {author} {\bibfnamefont {M.}~\bibnamefont
  {Belkin}}, \bibinfo {author} {\bibfnamefont {Q.~J.}\ \bibnamefont {Wang}},
  \bibinfo {author} {\bibfnamefont {C.}~\bibnamefont {Pflugl}}, \bibinfo
  {author} {\bibfnamefont {A.}~\bibnamefont {Belyanin}}, \bibinfo {author}
  {\bibfnamefont {S.}~\bibnamefont {Khanna}}, \bibinfo {author} {\bibfnamefont
  {A.}~\bibnamefont {Davies}}, \bibinfo {author} {\bibfnamefont
  {E.}~\bibnamefont {Linfield}}, \ and\ \bibinfo {author} {\bibfnamefont
  {F.}~\bibnamefont {Capasso}},\ }\href {\doibase 10.1109/JSTQE.2009.2013183}
  {\bibfield  {journal} {\bibinfo  {journal} {Selected Topics in Quantum
  Electronics, IEEE Journal of}\ }\textbf {\bibinfo {volume} {15}},\ \bibinfo
  {pages} {952} (\bibinfo {year} {2009})}\BibitemShut {NoStop}%
\bibitem [{\citenamefont {Williams}(2007)}]{Williams}%
  \BibitemOpen
  \bibfield  {author} {\bibinfo {author} {\bibfnamefont {B.~S.}\ \bibnamefont
  {Williams}},\ }\href@noop {} {\bibfield  {journal} {\bibinfo  {journal} {Nat
  Photon}\ }\textbf {\bibinfo {volume} {1}},\ \bibinfo {pages} {517} (\bibinfo
  {year} {2007})}\BibitemShut {NoStop}%
\bibitem [{\citenamefont {Ferreira}\ and\ \citenamefont
  {Bastard}(1989)}]{Ferreira}%
  \BibitemOpen
  \bibfield  {author} {\bibinfo {author} {\bibfnamefont {R.}~\bibnamefont
  {Ferreira}}\ and\ \bibinfo {author} {\bibfnamefont {G.}~\bibnamefont
  {Bastard}},\ }\href {\doibase 10.1103/PhysRevB.40.1074} {\bibfield  {journal}
  {\bibinfo  {journal} {Phys. Rev. B}\ }\textbf {\bibinfo {volume} {40}},\
  \bibinfo {pages} {1074} (\bibinfo {year} {1989})}\BibitemShut {NoStop}%
\bibitem [{\citenamefont {Smet}\ \emph {et~al.}(1996)\citenamefont {Smet},
  \citenamefont {Fonstad},\ and\ \citenamefont {Hu}}]{Smet}%
  \BibitemOpen
  \bibfield  {author} {\bibinfo {author} {\bibfnamefont {J.~H.}\ \bibnamefont
  {Smet}}, \bibinfo {author} {\bibfnamefont {C.~G.}\ \bibnamefont {Fonstad}}, \
  and\ \bibinfo {author} {\bibfnamefont {Q.}~\bibnamefont {Hu}},\ }\href@noop
  {} {\bibfield  {journal} {\bibinfo  {journal} {Journal of Applied Physics}\
  }\textbf {\bibinfo {volume} {79}} (\bibinfo {year} {1996})}\BibitemShut
  {NoStop}%
\bibitem [{\citenamefont {Bastard}(1988)}]{Bastard1988}%
  \BibitemOpen
  \bibfield  {author} {\bibinfo {author} {\bibfnamefont {G.}~\bibnamefont
  {Bastard}},\ }\enquote {\bibinfo {title} {Wave mechanics applied to
  semiconductor heterostructures},}\ \ (\bibinfo  {publisher} {Les {\'E}ditions
  de Physique},\ \bibinfo {year} {1988})\BibitemShut {NoStop}%
\bibitem [{\citenamefont {Smith}\ and\ \citenamefont
  {Mailhiot}(1990)}]{Smith1990}%
  \BibitemOpen
  \bibfield  {author} {\bibinfo {author} {\bibfnamefont {D.~L.}\ \bibnamefont
  {Smith}}\ and\ \bibinfo {author} {\bibfnamefont {C.}~\bibnamefont
  {Mailhiot}},\ }\href {\doibase 10.1103/RevModPhys.62.173} {\bibfield
  {journal} {\bibinfo  {journal} {Rev. Mod. Phys.}\ }\textbf {\bibinfo {volume}
  {62}},\ \bibinfo {pages} {173} (\bibinfo {year} {1990})}\BibitemShut
  {NoStop}%
\bibitem [{\citenamefont {Kim}\ \emph {et~al.}(2002)\citenamefont {Kim},
  \citenamefont {Kent}, \citenamefont {Zunger},\ and\ \citenamefont
  {Geller}}]{Kim2002}%
  \BibitemOpen
  \bibfield  {author} {\bibinfo {author} {\bibfnamefont {K.}~\bibnamefont
  {Kim}}, \bibinfo {author} {\bibfnamefont {P.~R.~C.}\ \bibnamefont {Kent}},
  \bibinfo {author} {\bibfnamefont {A.}~\bibnamefont {Zunger}}, \ and\ \bibinfo
  {author} {\bibfnamefont {C.~B.}\ \bibnamefont {Geller}},\ }\href {\doibase
  10.1103/PhysRevB.66.045208} {\bibfield  {journal} {\bibinfo  {journal} {Phys.
  Rev. B}\ }\textbf {\bibinfo {volume} {66}},\ \bibinfo {pages} {045208}
  (\bibinfo {year} {2002})}\BibitemShut {NoStop}%
\bibitem [{\citenamefont {Jancu}\ \emph
  {et~al.}(1998{\natexlab{a}})\citenamefont {Jancu}, \citenamefont {Scholz},
  \citenamefont {Beltram},\ and\ \citenamefont {Bassani}}]{Jancu1998}%
  \BibitemOpen
  \bibfield  {author} {\bibinfo {author} {\bibfnamefont {J.-M.}\ \bibnamefont
  {Jancu}}, \bibinfo {author} {\bibfnamefont {R.}~\bibnamefont {Scholz}},
  \bibinfo {author} {\bibfnamefont {F.}~\bibnamefont {Beltram}}, \ and\
  \bibinfo {author} {\bibfnamefont {F.}~\bibnamefont {Bassani}},\ }\href
  {\doibase 10.1103/PhysRevB.57.6493} {\bibfield  {journal} {\bibinfo
  {journal} {Phys. Rev. B}\ }\textbf {\bibinfo {volume} {57}},\ \bibinfo
  {pages} {6493} (\bibinfo {year} {1998}{\natexlab{a}})}\BibitemShut {NoStop}%
\bibitem [{\citenamefont {Jancu}\ \emph {et~al.}(2004)\citenamefont {Jancu},
  \citenamefont {Scholz}, \citenamefont {La~Rocca}, \citenamefont {de~Andrada~e
  Silva},\ and\ \citenamefont {Voisin}}]{Jancu2004}%
  \BibitemOpen
  \bibfield  {author} {\bibinfo {author} {\bibfnamefont {J.-M.}\ \bibnamefont
  {Jancu}}, \bibinfo {author} {\bibfnamefont {R.}~\bibnamefont {Scholz}},
  \bibinfo {author} {\bibfnamefont {G.~C.}\ \bibnamefont {La~Rocca}}, \bibinfo
  {author} {\bibfnamefont {E.~A.}\ \bibnamefont {de~Andrada~e Silva}}, \ and\
  \bibinfo {author} {\bibfnamefont {P.}~\bibnamefont {Voisin}},\ }\href
  {\doibase 10.1103/PhysRevB.70.121306} {\bibfield  {journal} {\bibinfo
  {journal} {Phys. Rev. B}\ }\textbf {\bibinfo {volume} {70}},\ \bibinfo
  {pages} {121306} (\bibinfo {year} {2004})}\BibitemShut {NoStop}%
\bibitem [{\citenamefont {Krebs}\ and\ \citenamefont
  {Voisin}(1996)}]{Krebs1996}%
  \BibitemOpen
  \bibfield  {author} {\bibinfo {author} {\bibfnamefont {O.}~\bibnamefont
  {Krebs}}\ and\ \bibinfo {author} {\bibfnamefont {P.}~\bibnamefont {Voisin}},\
  }\href {\doibase 10.1103/PhysRevLett.77.1829} {\bibfield  {journal} {\bibinfo
   {journal} {Phys. Rev. Lett.}\ }\textbf {\bibinfo {volume} {77}},\ \bibinfo
  {pages} {1829} (\bibinfo {year} {1996})}\BibitemShut {NoStop}%
\bibitem [{\citenamefont {Toropov}\ \emph {et~al.}(2000)\citenamefont
  {Toropov}, \citenamefont {Ivchenko}, \citenamefont {Krebs}, \citenamefont
  {Cortez}, \citenamefont {Voisin},\ and\ \citenamefont
  {Gentner}}]{Toropov2000}%
  \BibitemOpen
  \bibfield  {author} {\bibinfo {author} {\bibfnamefont {A.~A.}\ \bibnamefont
  {Toropov}}, \bibinfo {author} {\bibfnamefont {E.~L.}\ \bibnamefont
  {Ivchenko}}, \bibinfo {author} {\bibfnamefont {O.}~\bibnamefont {Krebs}},
  \bibinfo {author} {\bibfnamefont {S.}~\bibnamefont {Cortez}}, \bibinfo
  {author} {\bibfnamefont {P.}~\bibnamefont {Voisin}}, \ and\ \bibinfo {author}
  {\bibfnamefont {J.~L.}\ \bibnamefont {Gentner}},\ }\href {\doibase
  10.1103/PhysRevB.63.035302} {\bibfield  {journal} {\bibinfo  {journal} {Phys.
  Rev. B}\ }\textbf {\bibinfo {volume} {63}},\ \bibinfo {pages} {035302}
  (\bibinfo {year} {2000})}\BibitemShut {NoStop}%
\bibitem [{\citenamefont {Keating}(1966)}]{Keating}%
  \BibitemOpen
  \bibfield  {author} {\bibinfo {author} {\bibfnamefont {P.~N.}\ \bibnamefont
  {Keating}},\ }\href {\doibase 10.1103/PhysRev.145.637} {\bibfield  {journal}
  {\bibinfo  {journal} {Phys. Rev.}\ }\textbf {\bibinfo {volume} {145}},\
  \bibinfo {pages} {637} (\bibinfo {year} {1966})}\BibitemShut {NoStop}%
\bibitem [{\citenamefont {Brandt}\ \emph {et~al.}(1990)\citenamefont {Brandt},
  \citenamefont {Tapfer}, \citenamefont {Cingolani}, \citenamefont {Ploog},
  \citenamefont {Hohenstein},\ and\ \citenamefont {Phillipp}}]{Brandt1990}%
  \BibitemOpen
  \bibfield  {author} {\bibinfo {author} {\bibfnamefont {O.}~\bibnamefont
  {Brandt}}, \bibinfo {author} {\bibfnamefont {L.}~\bibnamefont {Tapfer}},
  \bibinfo {author} {\bibfnamefont {R.}~\bibnamefont {Cingolani}}, \bibinfo
  {author} {\bibfnamefont {K.}~\bibnamefont {Ploog}}, \bibinfo {author}
  {\bibfnamefont {M.}~\bibnamefont {Hohenstein}}, \ and\ \bibinfo {author}
  {\bibfnamefont {F.}~\bibnamefont {Phillipp}},\ }\href {\doibase
  10.1103/PhysRevB.41.12599} {\bibfield  {journal} {\bibinfo  {journal} {Phys.
  Rev. B}\ }\textbf {\bibinfo {volume} {41}},\ \bibinfo {pages} {12599}
  (\bibinfo {year} {1990})}\BibitemShut {NoStop}%
\bibitem [{\citenamefont {Raouafi}\ \emph {et~al.}(2016)\citenamefont
  {Raouafi}, \citenamefont {Benchamekh}, \citenamefont {Nestoklon},
  \citenamefont {Jancu},\ and\ \citenamefont {Voisin}}]{Raouafi2016}%
  \BibitemOpen
  \bibfield  {author} {\bibinfo {author} {\bibfnamefont {F.}~\bibnamefont
  {Raouafi}}, \bibinfo {author} {\bibfnamefont {R.}~\bibnamefont {Benchamekh}},
  \bibinfo {author} {\bibfnamefont {M.~O.}\ \bibnamefont {Nestoklon}}, \bibinfo
  {author} {\bibfnamefont {J.-M.}\ \bibnamefont {Jancu}}, \ and\ \bibinfo
  {author} {\bibfnamefont {P.}~\bibnamefont {Voisin}},\ }\href
  {http://stacks.iop.org/0953-8984/28/i=4/a=045001} {\bibfield  {journal}
  {\bibinfo  {journal} {Journal of Physics: Condensed Matter}\ }\textbf
  {\bibinfo {volume} {28}},\ \bibinfo {pages} {045001} (\bibinfo {year}
  {2016})}\BibitemShut {NoStop}%
\bibitem [{\citenamefont {Lew Yan~Voon}\ and\ \citenamefont
  {Ram-Mohan}(1993)}]{Ram-Mohan}%
  \BibitemOpen
  \bibfield  {author} {\bibinfo {author} {\bibfnamefont {L.~C.}\ \bibnamefont
  {Lew Yan~Voon}}\ and\ \bibinfo {author} {\bibfnamefont {L.~R.}\ \bibnamefont
  {Ram-Mohan}},\ }\href {\doibase 10.1103/PhysRevB.47.15500} {\bibfield
  {journal} {\bibinfo  {journal} {Phys. Rev. B}\ }\textbf {\bibinfo {volume}
  {47}},\ \bibinfo {pages} {15500} (\bibinfo {year} {1993})}\BibitemShut
  {NoStop}%
\bibitem [{\citenamefont {Krebs}\ \emph {et~al.}(1997)\citenamefont {Krebs},
  \citenamefont {Seidel}, \citenamefont {André}, \citenamefont {Bertho},
  \citenamefont {Jouanin},\ and\ \citenamefont {Voisin}}]{Krebs1997}%
  \BibitemOpen
  \bibfield  {author} {\bibinfo {author} {\bibfnamefont {O.}~\bibnamefont
  {Krebs}}, \bibinfo {author} {\bibfnamefont {W.}~\bibnamefont {Seidel}},
  \bibinfo {author} {\bibfnamefont {J.~P.}\ \bibnamefont {André}}, \bibinfo
  {author} {\bibfnamefont {D.}~\bibnamefont {Bertho}}, \bibinfo {author}
  {\bibfnamefont {C.}~\bibnamefont {Jouanin}}, \ and\ \bibinfo {author}
  {\bibfnamefont {P.}~\bibnamefont {Voisin}},\ }\href
  {http://stacks.iop.org/0268-1242/12/i=7/a=002} {\bibfield  {journal}
  {\bibinfo  {journal} {Semiconductor Science and Technology}\ }\textbf
  {\bibinfo {volume} {12}},\ \bibinfo {pages} {938} (\bibinfo {year}
  {1997})}\BibitemShut {NoStop}%
\bibitem [{\citenamefont {Jancu}\ \emph
  {et~al.}(1998{\natexlab{b}})\citenamefont {Jancu}, \citenamefont
  {Pellegrini}, \citenamefont {Colombelli}, \citenamefont {Beltram},
  \citenamefont {Mueller}, \citenamefont {Sorba},\ and\ \citenamefont
  {Franciosi}}]{Jancu98}%
  \BibitemOpen
  \bibfield  {author} {\bibinfo {author} {\bibfnamefont {J.~M.}\ \bibnamefont
  {Jancu}}, \bibinfo {author} {\bibfnamefont {V.}~\bibnamefont {Pellegrini}},
  \bibinfo {author} {\bibfnamefont {R.}~\bibnamefont {Colombelli}}, \bibinfo
  {author} {\bibfnamefont {F.}~\bibnamefont {Beltram}}, \bibinfo {author}
  {\bibfnamefont {B.}~\bibnamefont {Mueller}}, \bibinfo {author} {\bibfnamefont
  {L.}~\bibnamefont {Sorba}}, \ and\ \bibinfo {author} {\bibfnamefont
  {A.}~\bibnamefont {Franciosi}},\ }\href {\doibase
  http://dx.doi.org/10.1063/1.122525} {\bibfield  {journal} {\bibinfo
  {journal} {Applied Physics Letters}\ }\textbf {\bibinfo {volume} {73}},\
  \bibinfo {pages} {2621} (\bibinfo {year} {1998}{\natexlab{b}})}\BibitemShut
  {NoStop}%
\bibitem [{\citenamefont {Garcia}\ \emph {et~al.}(2000)\citenamefont {Garcia},
  \citenamefont {De~Nardis}, \citenamefont {Pellegrini}, \citenamefont {Jancu},
  \citenamefont {Beltram}, \citenamefont {Müeller}, \citenamefont {Sorba},\
  and\ \citenamefont {Franciosi}}]{Jancu2000}%
  \BibitemOpen
  \bibfield  {author} {\bibinfo {author} {\bibfnamefont {C.~P.}\ \bibnamefont
  {Garcia}}, \bibinfo {author} {\bibfnamefont {A.}~\bibnamefont {De~Nardis}},
  \bibinfo {author} {\bibfnamefont {V.}~\bibnamefont {Pellegrini}}, \bibinfo
  {author} {\bibfnamefont {J.~M.}\ \bibnamefont {Jancu}}, \bibinfo {author}
  {\bibfnamefont {F.}~\bibnamefont {Beltram}}, \bibinfo {author} {\bibfnamefont
  {B.~H.}\ \bibnamefont {Müeller}}, \bibinfo {author} {\bibfnamefont
  {L.}~\bibnamefont {Sorba}}, \ and\ \bibinfo {author} {\bibfnamefont
  {A.}~\bibnamefont {Franciosi}},\ }\href@noop {} {\bibfield  {journal}
  {\bibinfo  {journal} {Applied Physics Letters}\ }\textbf {\bibinfo {volume}
  {77}} (\bibinfo {year} {2000})}\BibitemShut {NoStop}%
\end{thebibliography}%

\bibliographystyle{apsrev4-1}

\end{document}